\documentclass[a4paper,11pt,onecolumn]{article}
\usepackage{latexsym}
\usepackage{dcolumn}
\usepackage{authblk}
\usepackage{blindtext}
\usepackage{hyperref}
\usepackage{graphicx}
\usepackage{braket,amsmath}
\usepackage{bm}       
\usepackage{amssymb}

\date{}
\begin{document}

\title{ Schwarzschild phase without a black hole}

\author{Sandipan Sengupta \\ Department of Physics and Centre for Theoretical Studies, \\Indian Institute of Technology Kharagpur, Kharagpur-721302, India\\
sandipan@phy.iitkgp.ac.in}
\maketitle

\begin{abstract}
We present a smooth extension of the Schwarzschild exterior geometry, where the singular interior is superceded by a vacuum phase with vanishing metric determinant.  Unlike  the  Kruskal-Szekeres  continuation, this solution to the first order field equations in vacuum has no singularity in the curvature two-form fields, no horizon and no global time. The underlying non-analytic structure provides a distinct geometric realization of `mass' in classical gravity. 
We also find that the negative mass Schwarzschild solution does not admit a similar extension within the first order theory. This is consistent with the general expectation that degenerate metric solutions associated with the Hilbert-Palatini Lagrangian formulation should satisfy the energy conditions.
\end{abstract}
\vspace{.2cm}

\maketitle


\section{Introduction}

In view of the imposing experimental success of Einstein's theory of general relativity, one could be tempted to accept the invertible metric phase ($\det g_{\mu\nu}\neq 0$) as a self-contained and complete framework in describing the classical dynamics of spacetimes.
However, in general, (the first order formulation of) gravity theory is known to exhibit an additional phase based on non-invertible metrics \cite{tseytlin,kaul1,kaul2}. From a general perspective, the Einsteinian theory is nothing more or less than the special phase of first order theory where the metric is invertible everywhere. It then seems natural to ask as to how certain robust features of Einsteinian solutions, such as singularities, get manifested in a more generic solution where both the phases could coexist.

In the spherically symmetric vacuum Einstein theory, the Schwarzschild spacetime happens to be the unique solution. For a positive (negative) mass, its singular interior represents a black hole with a horizon (a naked singularity without a horizon). However, in the presence of both the phases of first order gravity, there is no reason for Birkhoff's uniqueness theorem to be applicable, as it works only under the assumption of the invertibility of metric. 
Here we provide an explicit realization of such a scenario \cite{bengtsson,kaul}.
The solutions discussed are smooth extensions of the Schwarzschild exterior through a noninvertible metric phase, which supercedes the interior of the standard Schwarzschild solution of Einstein equations. These spacetimes are solutions to the first order equations of motion everywhere. The associated field-strength components are manifestly finite everywhere, unlike in the standard Schwarzschild case.  Even though it is not possible to construct four dimensional curvature scalars in regions where the metric is degenerate, it is still possible to define effective lower-dimensional curvature scalars associated with the nondegenerate subspace of the four geometry in such regions. All such scalars are found to be finite, implying that the emergent three-geometry is regular. 

Let us now elucidate our construction for the positive mass Schwarzschild geometry, followed by a brief discussion of the negative mass case.


\section{A smooth extension of the Schwarzschild exterior}
Our aim is to construct a smooth continuation of the Schwarzschild exterior, such that it satisfies the vacuum field equations of first order gravity\footnote{The form of the metric defining the full spacetime here is the same as introduced in ref.\cite{kaul}. However, the solutions there are based on fields that are continuous but not smooth and have nonvanishing torsion, unlike the case here.}. These equations of motion are obtained through a variation of the Hilbert-Palatini Lagrangian density with respect to the tetrad and connection fields:
\begin{eqnarray}\label{eom1}
e_{[\mu}^{[K} D_{\nu}(\omega)e_{\alpha]}^{L]}=0,~
e_{[\nu}^{[J} R_{\alpha\beta]}^{~KL]}(\omega)=0~.
\end{eqnarray}
Although degenerate solutions to these equations may be associated with nontrivial torsion in general, we shall consider torsionless configurations which would be sufficient for our purpose here.

\subsection{Basic fields:} We define the two phases of the full spacetime through the following metric $\left(t\in (-\infty,\infty),~u\in 
(-\infty,\infty),~\theta\in[0,\pi],~\phi\in[0,2\pi]\right)$:
\begin{eqnarray}\label{sch1}
ds^2 &=&-\left[1-\frac{2M}{f(u)}\right]dt^2 +
\frac{f'^{2}(u)}{\left[1-\frac{2M}{f(u)}\right]} du^2 + 
f^2(u)\left[d\theta^2+\sin^2 \theta d\phi^2\right] ~(u>u_0),\nonumber\\
&=& 0+\sigma F^2(u) du^2 + H^2(u)\left[d\theta^2+\sin^2 
\theta d\phi^2\right]~(u\leq u_0).
\end{eqnarray}
The smooth functions $f,F$ satisfy the following set of boundary conditions:
\begin{eqnarray}\label{fF2} 
&& f(u)\rightarrow 2M,~f'(u)\rightarrow 0,~F(u)\rightarrow 0~\mathrm{as~}u\rightarrow u_0;\nonumber\\
&& f(u)\rightarrow \infty,~f'(u)\rightarrow 1 ~\mathrm{as~}u\rightarrow \infty.
\end{eqnarray} 
While the first line above ensures the continuity of the metric, the last line implies that the spacetime is flat as $u\rightarrow \infty$.
The internal metric is given by $\eta_{IJ}=diag[-1,1,1,1]$.
The metric at $u>u_0$ may be brought to the Schwarzschild form through a reparametrization $u\rightarrow r=f(u)$. However, the degenerate metric at $u\leq u_0$ has no semblance to the Schwarzschild interior. Also, $u_0$ in this construction is not a new free parameter, but rather is dependent on the `mass' parameter M.
 Although it is not necessary to adopt any specific $f(u)$,  we choose it to be the following in order to be explicit:
\begin{eqnarray*}
f(u)=2M\left[1+\left|\frac{u}{u_0}-1\right| e^{-\frac{u_0^2}{(u-u_0)^2}}\right].
\end{eqnarray*}
This satisfies all the boundary conditions in (\ref{fF2}) provided $u_0=2M$.

The nonvanishing components of the associated (torsionless) spin-connection and the resulting field-strength are given below:
\begin{eqnarray}\label{omega}
\omega_t^{~01}&=&\frac{M}{f^2(u)},~\omega_\phi^{~23}=-\cos\theta,~
\omega_\phi^{~31}=\left(1-\frac{2M}{f(u)}\right)^{\frac{1}{2}}\sin\theta=-\omega_\theta^{~12}\sin\theta;\nonumber\\
 R_{tu}^{~01}&=&\frac{2Mf'(u)}{f^3(u)},~R_{t\phi}^{~03}=-\frac{M}
{f^2(u)}\left(1-\frac{2M}{f(u)}\right)^{\frac{1}{2}}
\sin\theta=R_{t\theta}^{~02}
\sin\theta ,\nonumber\\
R_{\theta\phi}^{~23}
&=& \frac{2M}{f(u)}\sin\theta,~
R_{\phi u}^{~31}=-\frac{Mf'(u)}{f^2(u)}\left(1-\frac{2M}{f(u)}
\right)^{-\frac{1}{2}}\sin\theta=R_{u\theta}^{~12}\sin\theta
\end{eqnarray}

 At the region $u\leq u_0$ with a degenerate phase, we choose a zero-torsion configuration, defined by the following connection fields along with the associated field-strength:
\begin{eqnarray}\label{R3}
\hat{\omega}_\phi^{~23}&=&-\cos\theta,~\hat{\omega}_\phi^{~31}
=\frac{H'(u)}{\sqrt{\sigma}F(u)}\sin\theta=-\hat{\omega}_\theta^{~12}\sin\theta;\nonumber\\
\hat{R}_{\theta\phi}^{~23}
&=&\left[1-\sigma\left(\frac{H'(u)}
{F(u)}\right)^2\right]\sin\theta,~\hat{R}_{\phi u}^{~31}=-\left[\frac{H'(u)}{\sqrt{\sigma}F(u)}\right]'\sin\theta=\hat{R}_{u\theta}^{~12}\sin\theta,
\end{eqnarray}
where we have displayed only the nonvanishing components. Note that even though some of the field  components are imaginary for $\sigma=-1$, the physical fields, given by their $SO(3,1)$ gauge-invariant counterparts $\hat{R}_{\mu\nu\alpha\beta}=\hat{R}_{\mu\nu}^{~~IJ} \hat{e}_{\mu I} \hat{e}_{\nu J}$, are all real.

\subsection{Solving the field equations:} Since this configuration with degenerate tetrads have vanishing torsion by construction, the first of the set of equations of motion (\ref{eom1}) is already satisfied. The remaining equation involving the curvature two-form is also satisfied provided the fields obey the constraint:
\begin{eqnarray}\label{mc}
\left(\frac{H'^2(u)}{F^2(u)}-\sigma\right)F(u)+2H(u)\left(\frac{H'(u)}{F(u)}\right)'=0
\end{eqnarray}
A solution to the continuity conditions at $u=u_0$ and the constraint (\ref{mc}) is given by:
\begin{eqnarray}
F(u)=-\frac{f'(u)}{\sqrt{\sigma}\left(1-\frac{2M}{f(u)}\right)^{\frac{1}{2}}},~H(u)=f(u)
\end{eqnarray}
Note that all the gauge covariant fields (tetrad and field-strength) are smooth across the phase boundary. The apparent discontinuity in the connection field ($\omega_t^{~01}\neq \hat{\omega}_t^{~01}$ at $u=u_0$) could be gauged away by a boost of the form:
\begin{eqnarray*}
\Lambda^I_{~J}=\left(\begin{array}{cccc}
\cosh \left[\frac{t}{4M}\right] & \sinh \left[\frac{t}{4M}\right] & 0 & 0\\
\sinh \left[\frac{t}{4M}\right] & \cosh \left[\frac{t}{4M}\right] & 0 & 0\\
0 & 0 & 1 & 0\\
 0 & 0 & 0 & 1  \end{array}\right)
\end{eqnarray*}
 
\subsection{Distinct features:}
We now summarize the main features of these new solutions of first order field equations:

a) The spacetime metric, along with all the gauge-covariant fields in this solution, are smooth everywhere.

b) The curvature two-form fields are finite everywhere, in contrast to the case of a Schwarzschild interior.

c) There is no horizon; Rather, the two-sphere at $u=u_0$ characterizes a minimal area surface $A_{min}=16\pi M^2$ and is impenetrable, at least classically.

d) The solution has a free parameter `M', exactly as in the usual Schwarzschild case. However, there is no matter sourcing this mass, rather, its origin is purely geometric.  
 
e) The method of defining the degenerate phase through $\hat{g}_{tt}=0$ is unique, since it is not possible to obtain a nontrivial extension of the Schwarzschild exterior through a degeneracy in any other direction (e.g. $\hat{g}_{uu}=0$).
 
Finally, note that even though the four-curvature scalar polynomials cannot be defined everywhere due to the noninvertibility of tetrads, the spacetime at $u\leq u_0$ may be treated as an emergent three-geometry defined solely by the induced metric:
\begin{eqnarray*} 
ds^2=\sigma F^2(u) du^2 + H^2(u)\left[d\theta^2+\sin^2 
\theta d\phi^2\right]
\end{eqnarray*}
The three-scalar curvature polynomials built upon the torsionless connection for this three-geometry also turn out to be finite everywhere. 


\section{Naked singularity}
Next, let us analyze the case of negative mass Schwarzschild geometry, the corresponding metric being obtained through a sign reversal $M\rightarrow -M$. This represents a naked singularity solution of vacuum Einstein equations.

Since there is no horizon in the original geometry $(g_{tt}\neq 0)$, let us consider an extension in first order gravity through $g_{uu}=0$ at $u\leq u_0$ is as follows:
\begin{eqnarray}
ds^2 &=&-\left[1+\frac{2M}{f(u)}\right]dt^2 +
\frac{f'^{2}(u)}{\left[1+\frac{2M}{f(u)}\right]} du^2 + 
f^2(u)\left[d\theta^2+\sin^2 \theta d\phi^2\right] ~(u>u_0),\nonumber\\
&=&-\left[1+\frac{2M}{f_0}\right]dt^2+ 0 + f^2_0\left[d\theta^2+\sin^2 
\theta d\phi^2\right]~(u\leq u_0),
\end{eqnarray}
where $u=u_0$ corresponds to the phase boundary hypersurface and $f_0=f(u_0)$. Since the phase boundary must correspond to some finite but nonzero Schwarzschild radius $(r)$, $f_0>0$. However, the noninvertible phase above just corresponds to the trivial restriction of the original negative mass solution at $u=u_0$ rather than a new spacetime region. The same conclusion holds for a possible degeneracy through $g_{tt}=0$ (based on a redefined `time').

Thus, the singular curvature-two form fields associated with `negative mass' solution cannot be regularized using a degenerate extension, in contrast to the `positive mass' case discussed in the previous section. 

\section {Conclusions}
The smooth extension of the exterior Schwarzschild geometry, as discussed here, involves a modification of the standard picture of a singular black hole interior in a purely classical setting \cite{kaul,sengupta}.  These solutions could provide a fresh perspective into the information loss problem. Being rooted within a fairly conservative framework based on first order gravity, such an approach could complement (and perhaps supercede) some of the more exotic proposals (e.g. Firewall or Fuzzball programme) that have been put forth to resolve this celebrated paradox.

The cases of positive and negative mass curvature singularities of the Einsteinian theory are perceived very differently within this framework. While the singular interior of positive mass Schwarzschild geometry may be traded for a zero-determinant phase with regular field-strength components, the negative mass naked singularity admits no such smooth extension in first order gravity. This outcome, however, is  consistent with the general expectation \cite{sam} that degenerate metric ($\det g_{\mu\nu}=0$) solutions obtained within the Hilbert-Palatini Lagrangian formulation \cite{tseytlin,kaul1,kaul2} should satisfy the energy conditions. The scenario here may be contrasted with the case of degenerate triad ($\det E^a_i=0$) solutions obtained earlier within the complex SU(2) (Sen-Ashtekar) Hamiltonian framework, which are known to contain negative energy geometries \cite{madhavan}.

Finally, let us note that in the vacuum solution constructed here, there is no matter sourcing the `mass' M. Rather, its origin could be attributed to the time-nonorientability at the phase boundaries \cite{kaul,sengupta,sengupta1,gera}. This is an interesting realization of mass through pure geometry. Further, this scenario is distinct from the remarkable ideas explored first by Einstein-Rosen \cite{einstein} and later by Wheeler-Misner \cite{wheeler}, through their respective attempts to generate `mass' and `charge' from a nontrivial geometry or topology. To emphasize, our analysis here does not involve wormholes (which are not solutions to the vacuum field equations in general) or quantum configurations such as geons.  It would be natural to ask if there is a more interesting manifestation of such geometric `mass' sourced by noninvertible phases in vacuum. 

To conclude, given the number of intriguing features of these `two-phased' solutions, one wonders if quantum gravity could be a more suitable arena for further explorations based on these.

\vskip1.5cm

\noindent{\bf Acknowledgements}

\vspace{0.5cm}
 The support of (in part) the ECR award grant no. ECR/2016/000027 under the SERB, DST, Govt. of India and of (in part) the ISIRD grant `RAQ' is gratefully acknowledged.


\vskip1cm

\begin{thebibliography}{0}

\bibitem{tseytlin} A. A. Tseytlin, {\em J. Phys.\ A: Math. Gen.} {\bf 15}, L105 (1982).

\bibitem{kaul1}  R.K. Kaul and S. Sengupta, {\em Phys. Rev.\ D} {\bf 93}, 084026 (2016).

\bibitem{kaul2} R.K. Kaul and S. Sengupta, {\em Phys. Rev.\ D} {\bf 94}, 104047 (2016).

\bibitem{bengtsson} I. Bengtsson, Class. Quantum Grav. 8 (1991) 1847-1858.

\bibitem{kaul} R.K. Kaul and S. Sengupta, {\em Phys. Rev.\ D} {\bf 96}, 104011 (2017).



\bibitem{sengupta} S. Sengupta, {\em Phys. Rev.\ D} {\bf 96}, 104031 (2017).

\bibitem{sam} J. Samuel, {\em Proc. Conf. of Physics at the Planck Scale (Puri), ed. J. Maharana} (1994).

\bibitem{madhavan} M. Varadarajan, {\em Class. Quantum Grav.} {\bf 8}, L235 (1991).

\bibitem{sengupta1} S. Sengupta, {\em Phys. Rev.\ D} {\bf 97}, 124038 (2018). 

\bibitem{gera} S. Gera and S. Sengupta, {\em Phys. Rev.\ D} {\bf 99}, 124038 (2019).

\bibitem{einstein} A. Einstein and N. Rosen, {\em Phys. Rev.} {\bf 48}, 73 (1935).

\bibitem{wheeler} J. A. Wheeler, {\em Phys. Rev.} {\bf 97}, 511, (1955);\\
C. W. Misner and J. A. Wheeler, {\em Annals Phys.} {\bf 2}, 525 (1957);\\
J. A. Wheeler, {\em Annals Phys.} {\bf 2}, 604 (1957);\\
R. Sorkin, {\em J. Phys. A} {\bf 10}, 717 (1977).








\end{thebibliography}
\end{document}